\documentstyle[aps,prd,floats]{revtex}
\begin{document}
\draft
\def\lsim{\lower.5ex\hbox{$\; \buildrel < \over \sim \;$}}
\def\gsim{\lower.5ex\hbox{$\; \buildrel > \over \sim \;$}}
\title{Behaviour of spin-$\frac{1}{2}$ particle around a charged black hole}
\author  {Banibrata Mukhopadhyay}

\vskip1.0cm
\address{Theoretical Astrophysics Group\\
S. N. Bose National Centre For Basic Sciences,\\
JD Block, Salt Lake, Sector-III, Calcutta-700091\\ 	
e-mail: bm@boson.bose.res.in  \\}
\baselineskip = 16 true pt
\maketitle
\vskip1.0cm
\setcounter{page}{1}
\noindent{Classical and Quantum Gravity}\\
\def\ch{\lower-0.55ex\hbox{--}\kern-0.55em{\lower0.15ex\hbox{$h$}}}
\def\lh{\lower-0.55ex\hbox{--}\kern-0.55em{\lower0.15ex\hbox{$\lambda$}}}	
{\it Received: 20th January, 2000; Accepted: 13th March, 2000}\\

\begin{abstract}

\baselineskip = 16 true pt

Dirac equation is separable in curved space-time and its solution
was found for both spherically and axially symmetric geometry. 
But most of the works were done without considering the charge
of the black hole. Here we consider the spherically symmetric
charged black hole background namely Reissner-Nordstr\"om black hole.
Due to presence of the charge of black-hole charge-charge interaction
will be important for the cases of incoming charged particle
(e.g. electron, proton etc.). Therefore both gravitational and
electromagnetic gauge fields should be introduced. Naturally behaviour 
of the particle will be changed from that in Schwarzschild geometry. 
We compare both the solutions. In the case of Reissner-N\"ordstrom
black hole there is a possibility of super-radiance unlike 
Schwarzschild case. We also check this branch of the solution. 

\end{abstract}

\pacs {04.20.-q, 04.70.-s, 04.70.Dy, 95.30.Sf}

\section*{I. INTRODUCTION}

Chandrasekhar separated Dirac equation in Kerr geometry into radial
and angular parts [1] in 1976. His separation method can be extended 
to Schwarzschild geometry and corresponding separated equations can
be found. But he did not consider the charge of the
black hole. If we consider the black hole as a charged one then 
electromagnetic interaction is important for incoming particle with charge. 
To study the behaviour of spin-$\frac{1}{2}$ particle,
Dirac wave is treated as a perturbation in the space-time
which is asymptotically flat [1]. Far away from the black hole
its influence on particle is not significant. As it comes closer,
feels the curvature of the space-time and corresponding behaviours
start to change with respect to that of flat space. Their behaviour around
black-hole without charge have been studied in the past by several authors [1-6].
In this paper, we will introduce charge in the black hole.
Here, we study a simpler problem to have a feeling about the
solution when the black hole is non-rotating but charged.
Here we have to solve Dirac equation in electromagnetic
field around a Reissner-Nordstr\"om black-hole. Thus we will
study the particle in {\it crossed} electromagnetic and gravitational field.
It is very clear that the potential felt by the incoming
Dirac wave will be different from that for Schwarzschild black hole [5].
For the incoming uncharged particle like neutron, electromagnetic
field does not play any part and the Dirac equation will be reduced to same as
Schwarzschild case except the re-definition of horizon. For charged incoming 
particle like electron, proton etc. electromagnetic gauge field should be
introduced. One also can study the neutrino wave whose behaviour 
is known for Kerr geometry [7].  In the next Section, we present the 
basic Dirac equations and separate them
in this crossed field. In \S 3, we study the behaviour of the potential 
and possibilities of super-radiance. In \S 4, we present a complete solution. 
Finally, in \S 5, we draw our conclusions.

\section*{II. DIRAC EQUATION AND IT'S SEPARATION}

By introducing electromagnetic interaction and gravitational effect
the covariant derivatives take the form as
$$
D_\mu=\partial_\mu+iq_1A_\mu+q_2\Gamma_\mu.
\eqno{(1)}
$$
The derivative of spinor $P^A$ can be written as
$$
D_\mu{P^A}=\partial_{\mu}P^A+iq_1A_\mu{P^A}+q_2\Gamma_{\mu\nu}^A{P^\nu},
\eqno{(2)}
$$
where, $q_1$ and $q_2$ are coupling constants. $q_1$ is the
charge of the incoming particle (say $q_1=q$) and $q_2$ is chosen throughout $1$.
$A_\mu$ and $\Gamma_{\mu\nu}^A$ are electromagnetic and gravitational
gauge (spin coefficients) fields respectively.
Thus, following [7] the Dirac equation in Newman-Penrose formalism 
can be written as
$$
\sigma_{AB'}^\mu{D_\mu}P^A+i\mu_p{\bar Q}^{C'}\epsilon_{C'B'}=0,
\eqno{(3a)}
$$
$$
\sigma_{AB'}^\mu{D_\mu}Q^A+i\mu_p{\bar P}^{C'}\epsilon_{C'B'}=0,
\eqno{(3b)}
$$
where, for any vector $X_i$, according to the spinor formalism [7] 
$\sigma_{AB'}^iX_i=X_{AB'}$; $A,B=0,1$. 
Here, we introduce a null tetrad $(\vec{l}, \vec{n}, \vec{m}, \vec{\bar{m}})$
to satisfy orthogonality relations, $\vec{l}{\bf .}\vec{n}=1$, 
$\vec{m}{\bf .}\vec{\bar{m}}=-1$ and $\vec{l}{\bf .}\vec{m}=\vec{n}{\bf .}\vec{m}=
\vec{l}{\bf .}\vec{\bar{m}}=\vec{n}{\bf .}\vec{\bar{m}}=0$ following 
Newman \& Penrose [8]. $2^{\frac{1}{2}}\mu_p$ is the mass of the Dirac particle. 
In terms of this new basis in Newman-Penrose formalism Pauli matrices can be written as
$$
\sigma_{AB'}^\mu=\frac{1}{\sqrt{2}}\left(
\begin{array}{cr} l^{\mu} & m^{\mu}\\{\bar{m}}^{\mu} & n^{\mu}\end{array}\right).
\eqno{(4)}
$$
Using equation (2), (3a), (4) and choosing $B=0$ and subsequently $B=1$ we get
$$
l^{\mu}(\partial_{\mu}+iqA_\mu)P^0+{\bar m}^{\mu}(\partial_\mu+iqA_\mu)P^1+
(\Gamma_{1000'}-\Gamma_{0010'}){P^0}+(\Gamma_{1100'}-\Gamma_{0110'}){P^1}-
i{\mu_p}{\bar Q}^{1'}=0,
\eqno{(5a)}
$$
$$
m^{\mu}(\partial_{\mu}+iqA_\mu)P^0+ n^{\mu}(\partial_\mu+iqA_\mu)P^1+
(\Gamma_{1001'}-\Gamma_{0011'}){P^0}+(\Gamma_{1101'}-\Gamma_{0111'}){P^1}+
i{\mu_p}{\bar Q}^{0'}=0,
\eqno{(5b)}
$$
Next by taking complex conjugation of equation (3b), writing various spin
coefficients by their named symbol [7] and choosing 
$$
P^0=F_1, P^1=F_2, {\bar Q}^{1'}=G_1, {\bar Q}^{0}=-G_2
$$ 
we get
$$
l^{\mu}(\partial_{\mu}+iqA_\mu)F_1+{\bar m}^{\mu}(\partial_\mu+iqA_\mu)F_2+
(\epsilon-\rho){F_1}+(\pi-\alpha){F_2}=i{\mu_p}G_1,
\eqno{(6a)}
$$
$$
m^{\mu}(\partial_{\mu}+iqA_\mu)F_1+n^{\mu}(\partial_\mu+iqA_\mu)F_2+
(\mu-\gamma){F_2}+(\beta-\tau){F_1}=i{\mu_p}G_2,
\eqno{(6b)}
$$
$$
l^{\mu}(\partial_{\mu}+iqA_\mu)G_2-m^{\mu}(\partial_\mu+iqA_\mu)G_1+
({\epsilon}^*-{\rho}^*){G_2}-({\pi}^*-{\alpha}^*){G_1}=i{\mu_p}F_2,
\eqno{(6c)}
$$
$$
n^{\mu}(\partial_{\mu}+iqA_\mu)G_1-{\bar m}^{\mu}(\partial_\mu+iqA_\mu)G_2+
({\mu}^*-{\gamma}^*){G_1}-({\beta}^*-{\tau}^*){G_2}=i{\mu_p}F_1,
\eqno{(6d)}
$$
These are the Dirac equations in Newman-Penrose formalism in curved
space-time with the presence of electromagnetic interaction.

Now we write the basis vectors of null tetrad in terms of elements of the Reissner-Nordstr\"om 
geometry [7,9] as,
$$
l^{\mu}=\frac{1}{\Delta}(r^2, \Delta, 0, 0),
\eqno{(7a)}
$$
$$
n^{\mu}=\frac{1}{2r^2}(r^2, -\Delta, 0, 0),
\eqno{(7b)}
$$
$$
m^{\mu}=\frac{1}{r\sqrt{2}}(0, 0, 1, icosec{\theta}),
\eqno{(7c)}
$$
$$
{\bar m}^{\mu}=\frac{1}{r\sqrt{2}}(0, 0, 1, -icosec{\theta}),
\eqno{(7d)}
$$
where, $\Delta=r^2-2Mr+Q_*^2$ and $G=\ch=c=1$ are chosen. Here $M$ is
mass of the black hole, $Q_*$ is charge of the black hole, $G$ is 
gravitational constant, $h$ is Plank's constant, $c$ is speed of light.

We consider the spin-$\frac{1}{2}$ wave function as the form of 
$e^{i({\sigma}t+m{\phi})}f(r,\theta)$ where, $\sigma$ is the frequency of
the incoming wave and $m$ is the azimuthal quantum number.
The temporal and azimuthal dependencies are chosen same but radial and polar
dependencies are chosen different for different spinors. Thus we write,
$$
f_1=e^{i({\sigma}t+m{\phi})}rF_1,\hskip0.2cm f_2=e^{i({\sigma}t+m{\phi})}F_2,\hskip0.2cm
g_1=e^{i({\sigma}t+m{\phi})}G_1, \hskip0.2cm g_2=e^{i({\sigma}t+m{\phi})}rG_2
\eqno{(8)}
$$ 
Now we strictly consider the static field so the magnetic potentials are
chosen zero, i.e., $A^{\mu}=(A^t,0,0,0)$. $A^t$ is nothing but corresponding
scaler potential of the field as (in this spherically symmetric space-time)
$$
A^t=\frac{qQ_{*}}{r-r_+},
\eqno{(9)}
$$
where, $r_+={\rm location\hskip0.1cm of\hskip0.1cm the\hskip0.1cm horizon}
=M+\sqrt{M^2-Q_*^2}$.

So using equations (7), (8) and (9) and writing various spin coefficients 
in terms of the Reissner-Nordstr\"om metric elements (actually in terms 
of basis vectors) [7] equation (6)s  reduce to
$$
{\cal D}_0f_1+2^{-\frac{1}{2}}{\cal L}_{\frac{1}{2}}f_2=i{\mu}_prg_1
\eqno{(10a)}
$$
$$
{\Delta}{\cal D}^{\dagger}_{\frac{1}{2}}f_2-2^{\frac{1}{2}}
{\cal L}_{\frac{1}{2}}^{\dagger}f_1=-2i{\mu}_prg_2
\eqno{(10b)}
$$
$$
{\cal D}_0g_2-2^{-\frac{1}{2}}{\cal L}_{\frac{1}{2}}^{\dagger}g_1=i{\mu}_prf_2
\eqno{(10c)}
$$
$$
{\Delta}{\cal D}^{\dagger}_{\frac{1}{2}}g_1+2^{\frac{1}{2}}
{\cal L}_{\frac{1}{2}}g_2=-2i{\mu}_prf_1
\eqno{(10d)}
$$
where, 
$$
{\cal D}_n=\frac{d}{dr}+\frac{ir^2\sigma}{\Delta}+\frac{iqQ_*r^2}{{\Delta}(r-r_+)}
+2n\frac{r-M}{\Delta}, \hskip0.8cm
{\cal D}_n^{\dagger}=\frac{d}{dr}-\frac{ir^2\sigma}{\Delta}
-\frac{iqQ_*r^2}{{\Delta}(r-r_+)}+2n\frac{r-M}{\Delta},
\eqno{(11)}
$$
$$
{\cal L}_{n}=\frac{d}{d\theta}+Q+ncot\theta,\hskip0.8cm
{\cal L}_n^{\dagger}=\frac{d}{d\theta}-Q+ncot\theta,\hskip0.8cm
Q=mcosec{\theta}.
\eqno{(12)}
$$
Now considering $f_1(r, \theta)=R_{-\frac{1}{2}}(r)S_{-\frac{1}{2}}(\theta)$, 
$f_2(r, \theta)=R_{\frac{1}{2}}(r)S_{\frac{1}{2}}(\theta)$, 
$g_1(r, \theta)=R_{\frac{1}{2}}(r)S_{-\frac{1}{2}}(\theta)$,
$g_2(r, \theta)=R_{-\frac{1}{2}}(r)S_{+\frac{1}{2}}(\theta)$ 
and following Chandrasekhar [7] we can separate the Dirac equation into radial and
angular parts as
$$
\Delta^{1/2}{\cal D}_0R_{-1/2}=(\lambda+im_pr){\Delta}^{1/2}R_{1/2},
\eqno{(13a)}
$$ 
$$
\Delta^{1/2}{\cal D}_0^{\dagger}{\Delta}^{1/2}R_{1/2}=(\lambda-im_pr)R_{-1/2},
\eqno{(13b)}
$$ 
$$
{\cal L}_{1/2}S_{1/2}=-{\lambda}S_{-1/2},
\eqno{(14a)}
$$
$$
{\cal L}_{1/2}^{\dagger}S_{-1/2}={\lambda}S_{1/2}.
\eqno{(14b)}
$$
Here, $m_p$ is the normalised rest mass of the incoming particle and $\lambda$ 
is the sepreration constant. 

\section*{III. Nature of the Potential in decoupled system}

The equations (14a-b) are same as the angular equation in Schwarzschild
geometry whose solution is given in [2,10,11] as
$$
\lambda^2=\left(l+\frac{1}{2}\right)^2, 
R_{\pm\frac{1}{2}}={\rm standared\hskip0.1cm spherical\hskip0.1cm harmonics}
=\hskip0.1cm   _{\pm \frac{1}{2}}\!Y^l_m(\theta).
\eqno{(15)}
$$
It is clear that the separation constant depends on orbital angular momentum
quantum number $l$. 

The equations (13a-b) are in coupled form. Following Chandrasekhar's [7] and
Mukhopadhyay \& Chakrabarti's [5] approach we can decouple it as
$$
\left(\frac{d^2}{d{\hat{r}}_*^2}+\sigma^2\right)Z_\pm={V_\pm}{Z_\pm},
\eqno{(16)}
$$
where, 
$$
{\hat{r}}_*=r_*+\frac{1}{2\sigma}tan^{-1}\frac{m_pr}{\lambda}+\frac{qQ_*}{\sigma}
\left[log(r-r_-)+\left\{\frac{2r_+}{r_+-r_-}-\frac{r_+^2}{(r_+-r_-)^2}\right\}
log\left(\frac{r-r_+}{r-r_-}\right)-\frac{r_+^2}{(r-r_-)(r-r_+)}\right],
\eqno{(17)}
$$
$$
r_*=r-3M+\frac{r_+^2}{r_+-r_-}log(r-r_+)-\frac{r_-^2}{r_+-r_-}log(r-r_-),
\eqno{(18)}
$$
$$
r_\pm=M\pm\sqrt{M^2-Q_*^2}, \hskip1cm
Z_\pm={\Delta}^{1/2}R_{1/2}e^{i{\Theta}/2}{\pm}R_{-1/2}e^{-i{\Theta}/2},
\hskip1cm\Theta=m_pr.
\eqno{(19)}
$$
In the extreme case when $M=Q^*$, expression for ${\hat{r}}_*$ and
$r_*$ are given as,
$$
{\hat{r}}_*=r_*+\frac{1}{2\sigma}tan^{-1}\frac{m_pr}{\lambda}+
\frac{qQ_*}{\sigma}\left[log(r-M)-\frac{2M^2}{(r-M)^2}-\frac{2M}{(r-M)}\right],
\eqno{(17')}
$$
$$
r_*=r-M+2Mlog(r-M)-\frac{M^2}{(r-M)}.
\eqno{(18')}
$$
Here, ${\hat{r}}_*$ is varying from $-\infty$ to $+\infty$ (cartesian
coordinate).
If we compare equation (16) with one dimensional Schr\"odinger equation 
in cartesian coordinate system the energy $E$ of the 
incoming particle can be written as $E{\propto}\sigma^2$ and the potential
($V_\pm$) felt by the particle is given as
$$
V_\pm=\frac{{\Delta}(\lambda^2+m_p^2r^2)^3}{\left[r^2(\lambda^2+m_p^2r^2)
\left(1+\frac{Q_*q}{(r-r_+)\sigma}\right)+\frac{{\Delta}{\lambda}m_p}{2\sigma}\right]^2}
\pm\frac{{\Delta}(\lambda^2+m_p^2r^2)}{\left[r^2(\lambda^2+m_p^2r^2)
\left(1+\frac{Q_*q}{(r-r_+)\sigma}\right)+\frac{{\Delta}{\lambda}m_p}{2\sigma}
\right]^3}
$$
$$
[\left\{r^2(\lambda^2+m_p^2r^2)\left(1+\frac{Q_*q}
{(r-r_+)\sigma}\right)+\frac{{\Delta}{\lambda}m_p}{2\sigma}\right\}
\frac{(\lambda^2+m_p^2r^2)^{1/2}}{\Delta^{1/2}}\{(r-M)(\lambda^2+m_p^2r^2)
+3{\Delta}rm_p^2\}-{\Delta}^{1/2}(\lambda^2+m_p^2r^2)^{3/2}
$$
$$
\left\{2r(\lambda^2 +m_p^2r^2)\left(1+\frac{Q_*q}{(r-r_+)\sigma}\right)
+2r^3m_p^2\left(1+\frac{Q_*q}{(r-r_+)\sigma}\right)
-r^2(\lambda^2+m_p^2r^2)\frac{Q_*q}
{(r-r_+)^2{\sigma}}+\frac{(r-M){\lambda}m_p}{\sigma}\right\}].
\eqno{(20)}
$$
From the expression of $V_\pm$ it is very clear that potential 
strictly depends on charge of the particle as well as of black hole.
More precisely it depends on Coulomb interaction between charge of black
hole and incoming particle. When charge of the black hole or particle
or both are chosen zero the potential reduces to same as that in Schwarzschild
geometry [5]. When factor $\frac{Q_*q}{\sigma}$ is positive 
potential varies smoothly. When $\frac{Q_*q}{\sigma}$ becomes negative 
$V_\pm$ diverges at a certain
location $r=\alpha$. For the second case factor $\left(1+\frac{Q_*q}
{(r-r_+)\sigma}\right)$ vanishes at $r=r_+-\frac{Q_*q}{\sigma}>r_+$ and
then becomes negative. At $r=\alpha>r_+$ denominator of $V_\pm$ vanishes.
For all other cases $\alpha<r_+$ always, so there is no scope to diverge
the potential. Thus for the positive energy solution when the 
electro-magnetic scalar potential in the field is of 
attractive nature corresponding potential diverges
again for negative energy solution potential diverges for repulsive 
electro-magnetic scalar potential. For the integral spin particle, 
it is found that
when potential diverges energy extraction is possible i.e., super-radiation is
occurred in the space-time [7]. On the other hand for the case of spin-half 
particle in Kerr geometry although at a certain parameter region potential
diverges but super-radiation does not exist [7]. In the case of spherically
symmetric Schwarzschild geometry potential does not diverge at all and
no scope of super-radiation [5]. Here it is interesting to note that
although our space-time is spherically symmetric but due to presence
of electromagnetic interaction term, the region exist which is expected 
to be super-radiant. 

Figure 1 shows behaviour of potential $V_+$ for
different values black hole charges, where $\sigma=0.8$, $m_p=0.8$,
$l=\frac{1}{2}$, $q=1$ are chosen; $\alpha<r_+$. 
When $Q_*=0$ (solid curve), potential reduces to
same as Schwarzschild case shown in Fig. 2 by Mukhopadhyay \& Chakrabarti [5].
It is also seen that with the increment of charge of the black hole, 
barrier height decreases. Increment of black hole charge indicates the
increment of electro-magnetic coupling and corresponding repulsive 
scalar potential opposes the attractive gravitational field. So net
effect decreases. Figure 2 shows the change of potential barrier for
different values of particle charge, where $\sigma=0.8$, $m_p=0.8$,
$l=\frac{1}{2}$, $Q_*=0.6$ are chosen; $\alpha<r_+$. Solid curve
indicates the potential felt by the neutron like particle. 

Now come to the cases when $\frac{Q_*q}{\sigma}$ is negative. 
For these cases ${\hat{r}}_*-r$ relation attends multivalues.
For both $r\rightarrow \infty$ and $r\rightarrow r_+$, 
$\hat{r}_* \rightarrow \infty$.
As explained above net potential barrier diverges at a certain location
in this parameter region.
From equation (20) it is very clear that near $r=\alpha$, potential
varies as $\frac{1}{(r-\alpha)^3}$. So it has two branches, one repulsive
and another attractive on each side of the singular point. As a result 
super-radiation is absent for the case of Reissner-Nordstr\"om geometry 
as in other cases [5,7]. We can choose any combination of 
$Q_*$, $q$ and $\sigma$ in such a way that $\frac{Q_*q}{\sigma}$ is negative.  

In Fig. 3 we show how nature of the potential ($V_+$) changes with rest 
mass of the incoming particle where, $\sigma=0.8$, $Q_*=0.5$, 
$l=\frac{1}{2}$, $q=1$ are chosen. Solid curve shows nature for neutrino wave. 
It is very clear from the figure that with the increase of rest mass of 
the incoming particle gravitational interaction increases and corresponding
potential barrier attains high value.

\section*{IV. The complete solution}

Now we will find spatially complete solution. As we mentioned
earlier that solution of the angular part is known which is same as
Schwarzschild case [5,10,11]. For radial solution we need to solve decoupled 
radial equation. The solution of equation (16) for potential $V_+$ and $V_-$,
using Instantaneous WKB Approximation (in short IWKB) method [5,6] can be written as  
$$
Z_{+} = \sqrt{T_+[k_+({\hat{r}_*})]} exp (i u_+) + \sqrt{R_+[k_+({\hat{r}_*})]} 
exp (- i u_+), 
\eqno{(21a)}
$$
$$
Z_{-} = \sqrt{T_-[k_-({\hat{r}_*})]} exp (i u_-) + \sqrt{R_-[k_-({\hat{r}_*})]} 
exp (- i u_-), 
\eqno{(21b)}
$$
where,
$$
k_{\pm} ({\hat r}_*) = \surd\left(\sigma^{2} - V_{\pm}\right),
\eqno{(22)}
$$
$$
u_{\pm} ({\hat r}_*) = \int k_{\pm}({\hat r}_*) d {\hat r}_* + {\rm constant},
\eqno{(23)}
$$
with
$$
T_+(r)+R_+(r)=1,\hskip0.5cm T_-(r)+R_-(r) = 1 \hskip0.5cm {\rm instantaneously}.
\eqno{(24)}
$$
Here, $k$ is the wavenumber of the incoming wave and $u$ is the {\it Eiconal},
$T_\pm$ and $R_\pm$ are instantaneous transmission and reflection coefficients
[5] respectively. Using this method at each location, instantaneously, WKB
method is applied. This solution is valid when ${1 \over k}
{dk \over d{\hat{r}}_{*}} << k$, otherwise different method [5] should be used.

In Fig. 4, the comparison of instantaneous reflection and transmission coefficients
in between Schwarzschild and Reissner-Nordstr\"om geometry are shown. The parameters
chosen are given in Figure Caption. With the decrease of barrier height the transmission
coefficient increases as well as reflection coefficient decreases. 
As it is seen that by introduction of the electromagnetic coupling, potential
barrier heights reduce so corresponding transmission probability
increase with respect to that of Schwarzschild case
(the behaviour for Schwarzschild case is graphically shown in [5])
for a particular set of parameter. So the presence of the charge of
black hole decreases the curved nature of space-time. 

Now recombining $Z_+$ and $Z_-$ one easily can find out original
radial Dirac wave functions $R_\frac{1}{2}$ and $R_{-\frac{1}{2}}$ [5].
Finally we will have complete solution as 
$J(r,\theta)={R_{\pm\frac{1}{2}}(r)}{S_{\pm\frac{1}{2}}(\theta)}$.

\section*{V. conclusions}

In this paper, we have studied analytically the scattering of spin-half
particles off Reissner-Nordstr\"om black hole. Our main motivation
is to show analytically how the spin-half particles behave in the presence of 
electromagnetic interaction in curved space-time. We introduced
the gravitational and electromagnetic gauge fields. Since no such kind of
study had been carried out previously we started from scratch.
Firstly, we wrote corresponding dynamical equation of spin-half
particle namely Dirac equation in combined gravitational and electromagnetic
background. Due to curvature of the space-time gravitational gauge field
(here, spin coefficients for Reissner-Nordstr\"om geometry) was introduced. The electromagnetic
interaction comes into the game because of charge of the black hole.
Here, we have considered steady-state problem and corresponding components of 
electromagnetic vector potential to zero. We then separated
the equation into radial and angular parts. It is seen that in case
of spherically symmetric space-time, presence of charge of the gravitating
object does not affect the behaviour of the incoming particles in polar
direction. Only the radial part of the equation is influenced. We then
decoupled the radial Dirac equation. Now the potential is dependent on
charge-charge coupling in the space-time. If the charge of the black hole
reduces to zero, the potential reduces to that of
Schwarzschild case. With the presence of repulsive (or attractive) 
charge-charge interaction for positive (or negative) energy solution 
magnitude of curvature effect reduces.  This is because of opposing
nature of two simultaneous interactions.     

There is one interesting sector of the solution (which sector was absent
in uncharged spherically symmetric space-time). If the charge-charge 
interaction is of attractive nature for positive energy solution (or
repulsive for negative energy solution) then potential at a certain
location ($r=\alpha$) diverges. But because of $\frac{1}{(r-\alpha)^3}$ variance of
potential super-radiation is absent.

Here we study the behaviour of potential by varying charge of the black hole,
charge of the incoming particle, rest mass of the incoming particle. We
also study the space-dependent reflection and transmission coefficients and
show graphically for one set of physical parameter. It is seen that as potential
barrier height decreases corresponding transmission probability increases. 
We solve the radial Dirac equation by IWKB method. 

\section*{Acknowledgment}

I am thankful to Prof. Sandip K. Chakrabarti regarding selection of this 
problem. I also like to thank to Kaushik Ghosh for helpful comments during my work.

\vskip 1cm



\section*{figure captions}

\noindent Fig. 1: Behaviour of potential for different values of the black hole charge.
Fixed parameters are, $\sigma=0.8$, $m_p=0.8$, $l=\frac{1}{2}$ and $q=1$. From
upper to lower curves the charge $Q_*$ of the black holes are chosen as 
$0,0.2,0.4,0.6,0.8,0.998$.  
\vskip0.5cm
\noindent Fig. 2: Behaviour of potential for different values of the incoming particle charge.
Fixed parameters are, $\sigma=0.8$, $m_p=0.8$, $l=\frac{1}{2}$ and $Q_*=0.6$. From
upper to lower curves the charge $q$ of the particles are chosen as 
$0,0.2,0.4,0.6,0.8,1$.  
\vskip0.5cm

\noindent Fig. 3: Behaviour of potential for different values of the rest mass of incoming particle.
Fixed parameters are, $\sigma=0.8$, $Q_*=0.5$, $l=\frac{1}{2}$ and $q=1$. From
upper to lower curves the mass $m_p$ of the particles are chosen as 
$0.4,0.3,0.2,0.1,0$.  
\vskip0.5cm

\noindent Fig. 4: Instantaneous reflection (R) and transmission (T) coefficients 
for Reissner-Nordstr\"om (solid curves) and Schwarzschild (dotted curves) black holes.
Physical parameters are chosen as $\sigma=0.8, m_p=0.8, 
l=\frac{1}{2}, q=1$. For Reissner-Nordstr\"om case $Q_*=0.5$.



\begin{references}

\bibitem {} Chandrasekhar, S. Proc. R. Soc. Lond. A {\bf 349}, 571 (1976).

\bibitem {} Chakrabarti, S.K. Proc. R. Soc. Lond. A {\bf 391}, 27 (1984).

\bibitem{} Jin, W.M. Class. Quantum Grav. {\bf 15}, 3163 (1998).

\bibitem{} Semiz, I. Phys. Rev. {\bf D46}, 5414 (1992).

\bibitem{} Mukhopadhyay, B. \& Chakrabarti S.K. Class. Quantum Grav. {\bf 16}, 3165 (1999). 

\bibitem{} Mukhopadhyay, B. Ind. J. Phys. {\bf 73B(6)}, 855 (1999). 

\bibitem {} Chandrasekhar, S. in {\it The Mathematical Theory Of Black Holes}
(London: Clarendon Press, 1983).

\bibitem{} Newman, E. \& Penrose, R. J. Math. Phys. {\bf 3}, 566 (1962).

\bibitem{} Kinnersley, W. Math. Phys. {\bf 10}, 1195 (1969).

\bibitem{} Newman, E. \& Penrose, R. J. Math. Phys. {\bf 7}, 863 (1966).

\bibitem{} Goldberg, J.N., Macfarlane, A.J., Newman, E.T., Rohrlich, F. \& Sudarsan, 
E.C.G. J. Math. Phys. {\bf 8}, 2155 (1967).

\end{references}
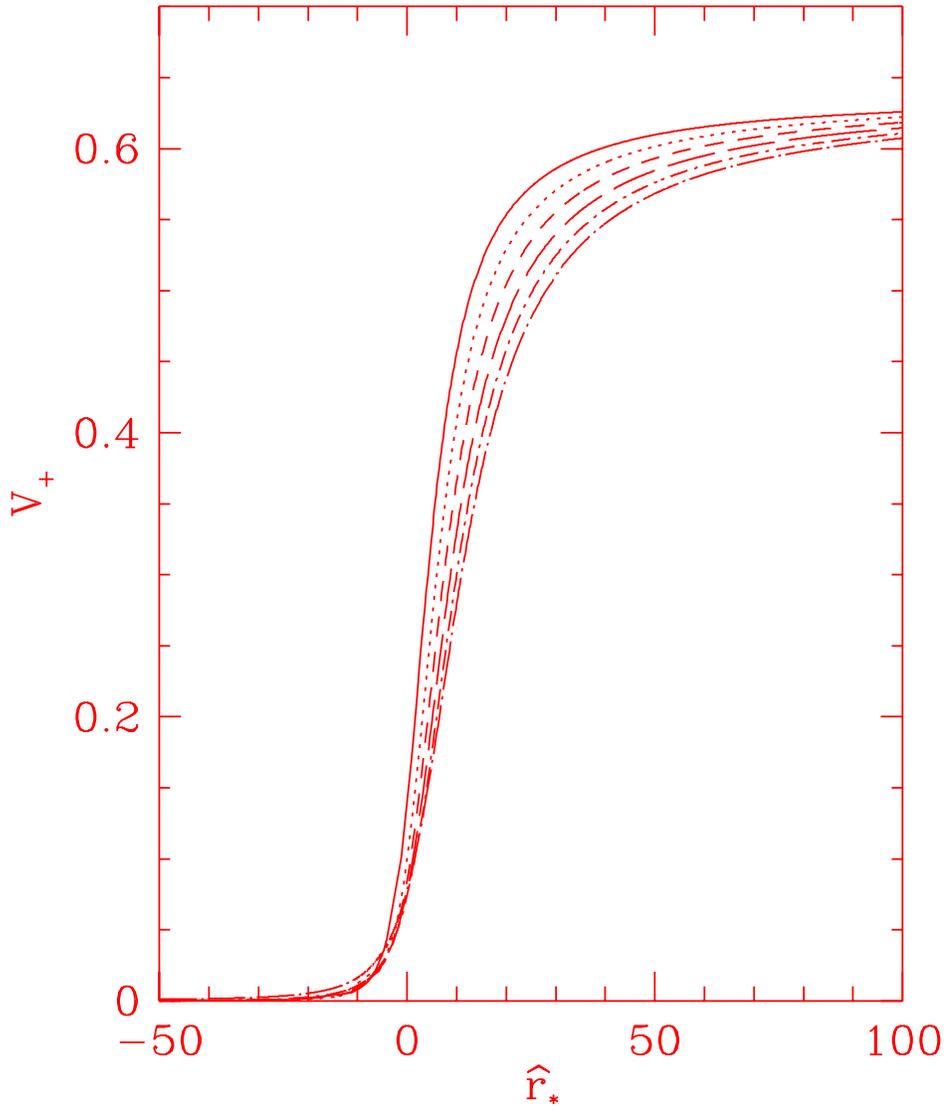
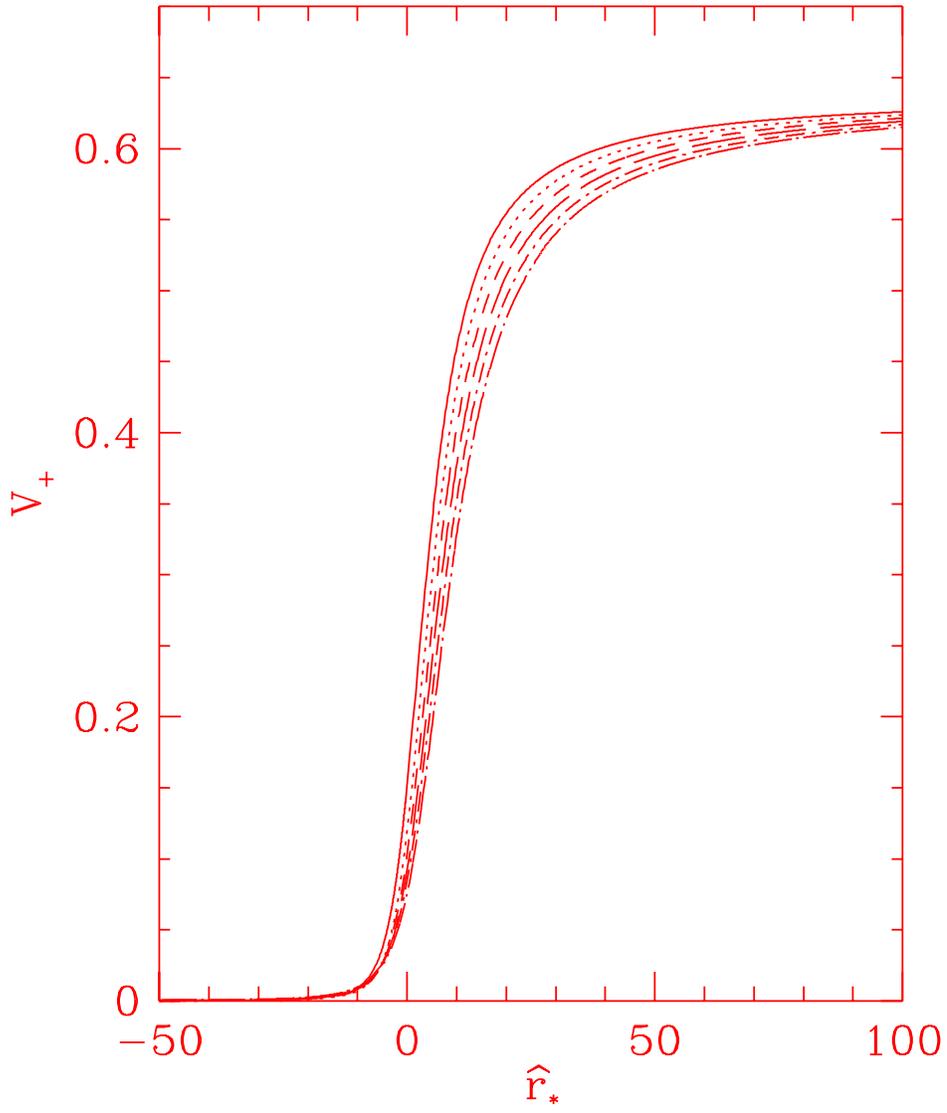
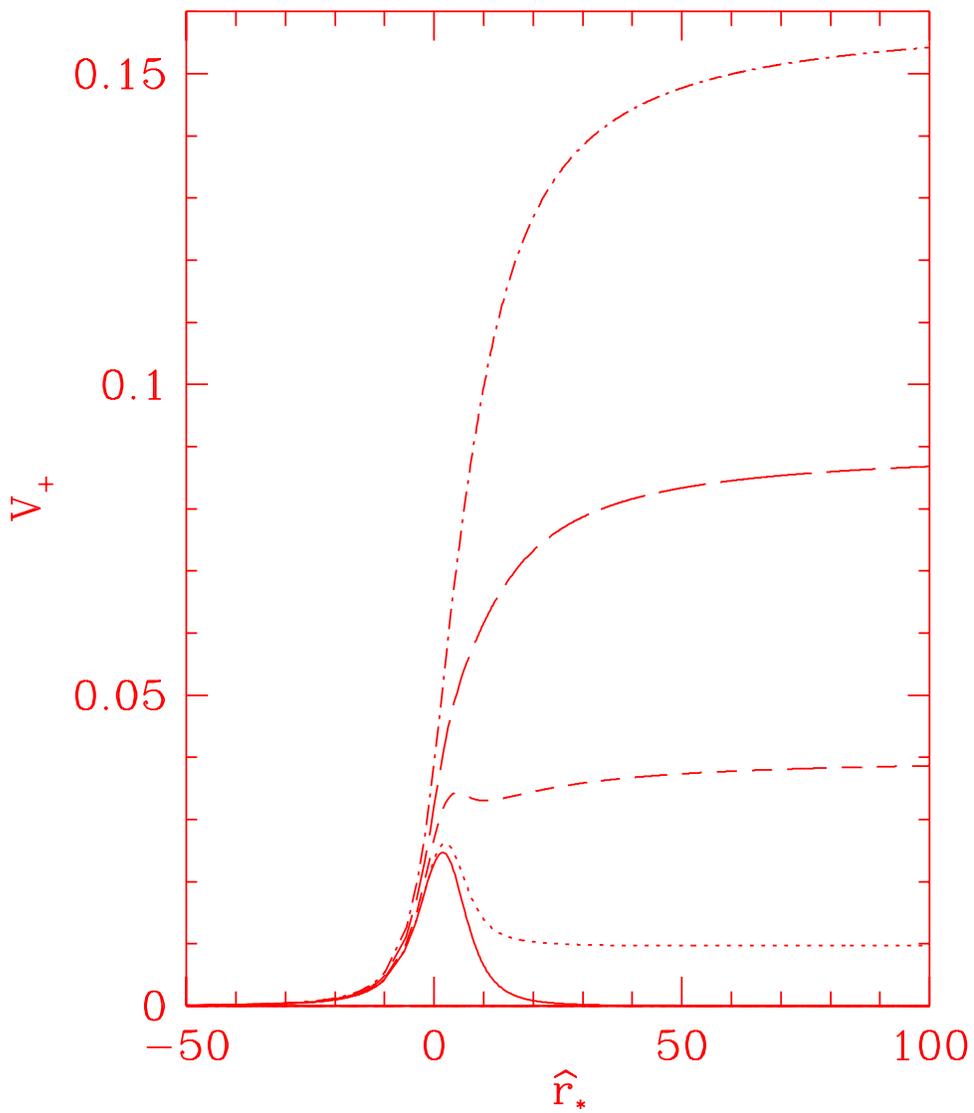
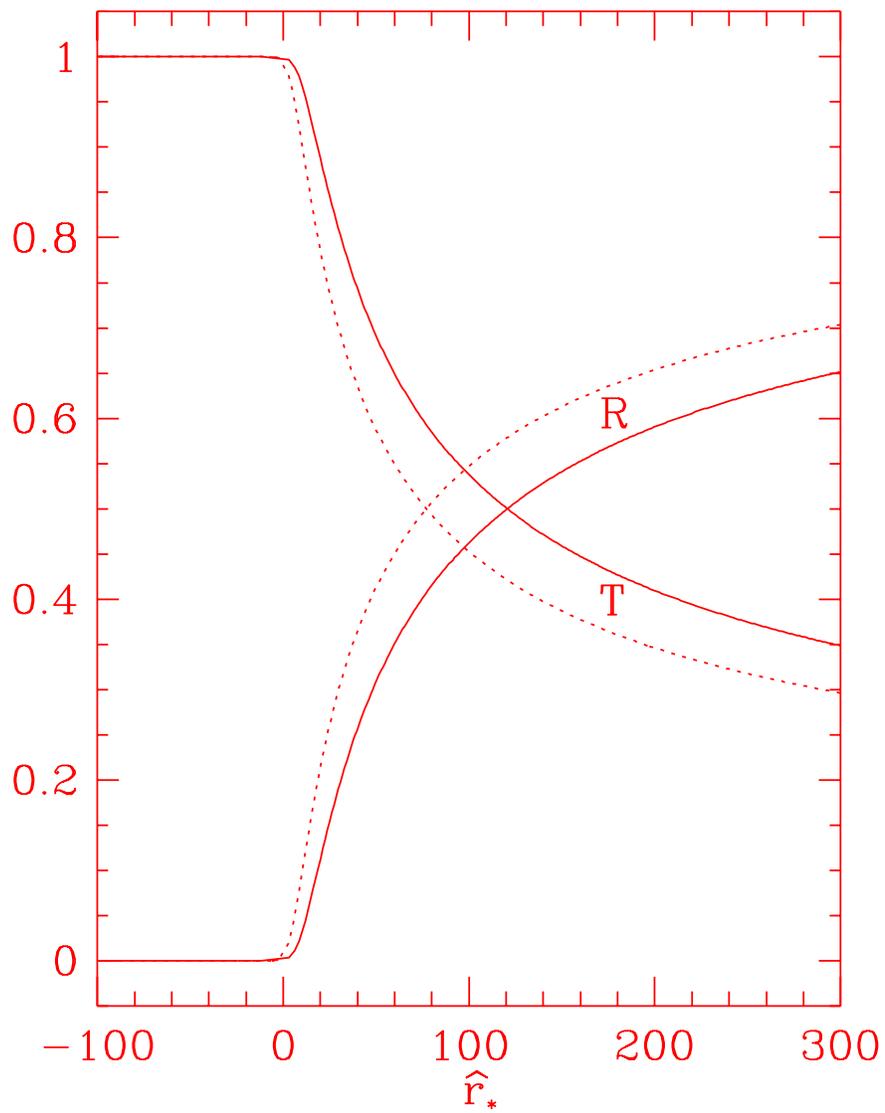
\end{document}